\documentclass[]{spie}  %>>> use for US letter paper
\usepackage[]{graphicx}
\usepackage[]{times}

\title{Commensal Observing with the Allen Telescope Array: Software Command and Control} 

\author{Colby Gutierrez-Kraybill\supit{a}, Garrett K. Keating\supit{a}, David MacMahon\supit{b}, Peter K. G. Williams\supit{b}\linebreak Gerald Harp\supit{c}, Robert Ackermann\supit{c}, Tom Kilsdonk\supit{c}, Jon Richards\supit{c}, William C. Barott\supit{d}
\skiplinehalf
\supit{a}University of California Berkeley, Hat Creek Radio Observatory\linebreak Bidwell Road, Hat Creek, CA USA 96040; \\
\supit{b}University of California Berkeley, Radio Astronomy Laboratory\linebreak 601 Campbell Hall, Berkeley, CA USA 97420; \\
\supit{c}SETI Institute, 515 N. Whisman Blvd, Mountain View, CA USA 94043; \\
\supit{d}Embry-Riddle University, Daytona Beach, FL USA 32114
}

\authorinfo{Further author information, email: colby@hcro.org telephone: 1 530 335 2364}
\begin{document} 
  \maketitle 

%%%%%%%%%%%%%%%%%%%%%%%%%%%%%%%%%%%%%%%%%%%%%%%%%%%%%%%%%%%%% 
\begin{abstract}
The Allen Telescope Array (ATA) is a Large-Number-Small-Diameter radio telescope array currently with 42 individual antennas and 5 independent back-end science systems (2 imaging FX correlators and 3 time domain beam formers) located at the Hat Creek Radio Observatory (HCRO).  The goal of the ATA is to run multiple back-ends simultaneously, supporting multiple science projects commensally.  The primary software control systems are based on a combination of Java, JRuby and Ruby on Rails.  The primary control API is simplified to provide easy integration with new back-end systems while the lower layers of the software stack are handled by a master observing system. Scheduling observations for the ATA is based on finding a union between the science needs of multiple projects and automatically determining an efficient path to operating the various sub-components to meet those needs.  When completed, the ATA is expected to be a world-class radio telescope, combining dedicated SETI projects with numerous radio astronomy science projects.
\end{abstract}

%>>>> Include a list of keywords after the abstract 


%%%%%%%%%%%%%%%%%%%%%%%%%%%%%%%%%%%%%%%%%%%%%%%%%%%%%%%%%%%%%
\section{INTRODUCTION}
\label{sec:intro}  % \label{} allows reference to this section
The Allen Telescope Array (ATA) is a Large-Number-Small-Diameter radio telescope array of 42 6-m antennas located at the Hat Creek Radio Observatory (HCRO) and will consist of 350 6-m antennas when completed.  Each antenna contains dual polarization receivers, providing continuous frequency coverage from 0.5-10 GHz. Four 600 MHz wide bands are made available from each antenna simultaneously at the back-end, which feed into one of 3 100 MHz wide spectral-imaging correlators or one of 3 time-domain beamformers.

The design goals for the ATA are specifically intended to enable commensal observing operations, allowing radio astronomy science activities while simultaneously feeding signals to other back-end systems, such as SETI signal detector systems, multi-beam sky-surveys and pulsar studies\cite{JWWelchIEEE09}.

In this paper, we describe an overview of the major components of the ATA, the software used to monitor and control those systems, and how observations logically divide the components between commensal projects.

Scheduling of commensal observations requires creating a union of requirements for various projects, competing for similar or identical time slots, sensitivity needs, \emph{uv}-plane coverage\cite{ATAmemo30} and the particular back-end(s) to fulfill a project's science goals.

%%%%%%%%%%%%%%%%%%%%%%%%%%%%%%%%%%%%%%%%%%%%%%%%%%%%%%%%%%%%%
\subsection{ATA Science Goals} 
\label{sec:title}  % \label{} allows reference to this section

The relatively wide field of view of the ATA antennas (2.45$^\circ$ at $\lambda$ = 21cm) naturally leads to large scale survey work dominating the key science goals of the ATA.  Many stars identified as SETI targets will be in every pointing direction of the array\cite{JWWelchIEEE09}.  Survey projects identified for the ATA are:

\begin{itemize}
\item Determine the HI content of galaxies out to z – 0.2 over 3π steradians, to measure how much intergalactic gas external galaxies are accreting; to search for dark, starless galaxies; to lay the foundation for Square Kilometer Array dark energy detection
\item Classify 250,000 extragalactic radio sources such as active galactic nuclei or starburst galaxies, to probe and quantify star formation in the local Universe; to identify high redshift objects; to probe large scale structure in the Universe; to identify gravitational lens candidates for dark matter and dark energy detection.
\item Explore the transient sky, to probe accretion onto black holes; to discover the origin of orphan gamma ray burst afterglows; to discover new and unknown transient phenomena.
\item Survey 1,000,000 stars for SETI emission with enough sensitivity to detect an Arecibo radar out to 300pc with in the frequency range of 1-10GHz.
\item Survey the 4x10$^{10}$ stars of the inner Galactic Plane from 1.42-1.72GHz for very powerful transmitters.
\item Measure the magnetic fields in the Milky Way and other Local group galaxies, to probe the role of magnetic fields in star formation and galaxy formation and evolution.
\item Detect the gravitational wave background from massive black holes through pulsar timing.
\item Measure molecular cloud and star formation properties using new molecular tracers, to map the star formation conditions on the scale of entire giant molecular clouds (GMCs); to determine the metallicity gradient of the Milky Way.
\end{itemize}

Regular commensal operations requires a careful orchestration of available resources for each project involved during a particular observation.

\section{SYSTEM DESCRIPTION}

\begin{figure}
\begin{center}
\begin{tabular}{c}
\includegraphics[height=11cm]{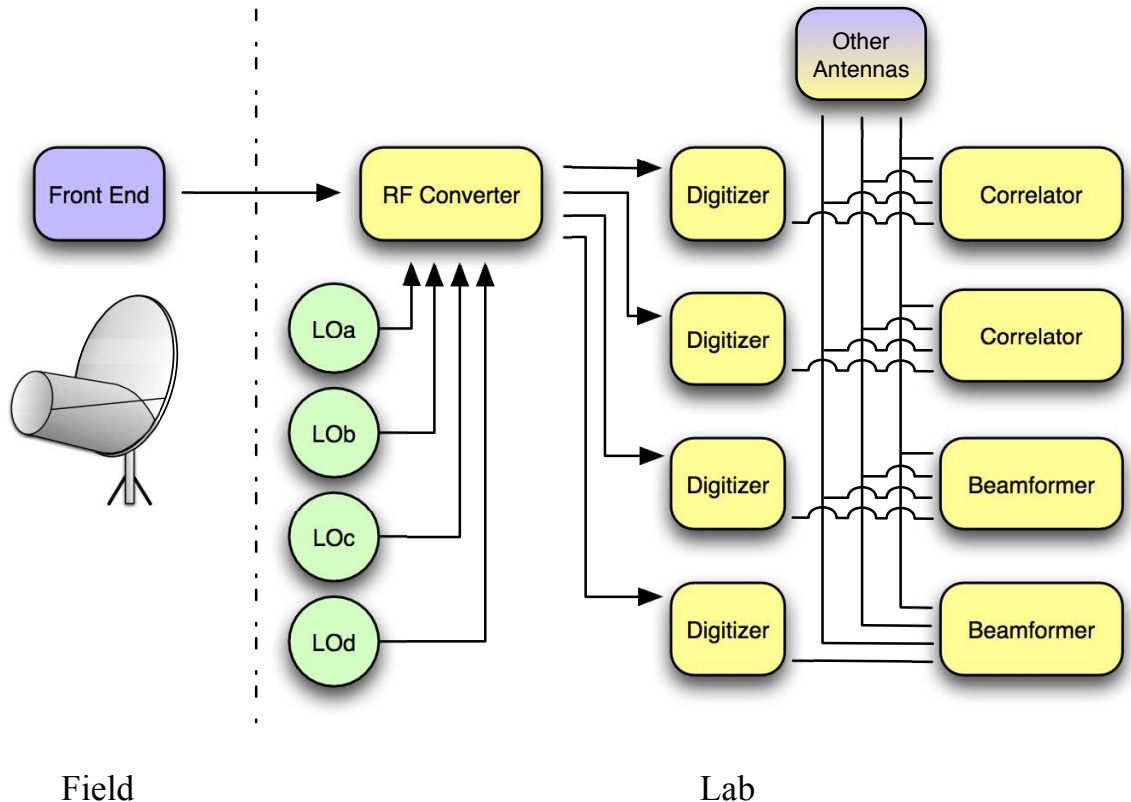}
\end{tabular}
\end{center}
\caption[ATA System Diagram] 
%>>>> use \label inside caption to get Fig. number with \ref{}
   { \label{fig:systemdiagram} 
ATA System diagram.}
\end{figure} 

The ATA System (see figure $\ref{fig:systemdiagram}$) is composed of front-end components; antenna assemblies, including drives, receiver control and monitoring systems and back-end components; Local Oscillators (LO's a - b), correlators and beamformers.

Each of these systems is controlled by one of two software libraries; Java based distributed applications and Ruby\cite{ruby} based scripts and a configuration database handled by Ruby on Rails\cite{ror}.

JRuby\cite{jruby} is utilized to create unified access to both Ruby and Java services.

\subsection{Java Simple Distributed Applications (JSDA)}

JSDA is a custom written light-weight remote method invocation protocol.  This library provides a subset of the services provided by CORBA\cite{omgcorba}, specifically, automatic generation of remote calling stubs and message routing.

\begin{figure}
\begin{center}
\begin{tabular}{c}
\includegraphics[height=9cm]{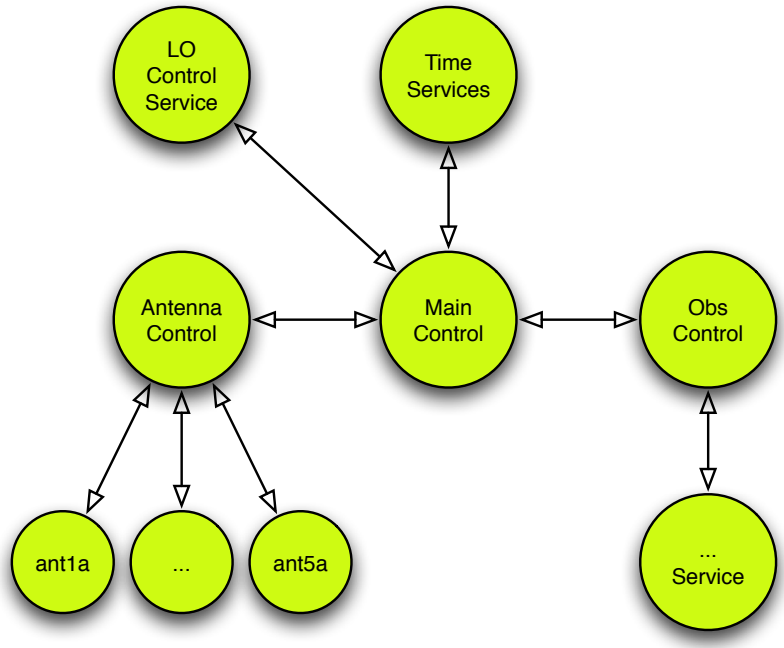}
\end{tabular}
\end{center}
\caption[ATA System Diagram] 
%>>>> use \label inside caption to get Fig. number with \ref{}
   { \label{fig:caterpillargraph} 
The JSDA network for the ATA, a Caterpillar Graph.  All nodes attaching to the JSDA network are within distance 1 of a central communications path.}
\end{figure}

The communications network provided by JSDA is a dynamically generated message bus, passing the necessary calling arguments for remote method invocation and transporting returned objects from java methods.  In particular, the configuration of the JSDA network for the ATA can be considered a caterpillar graph, see figure $\ref{fig:caterpillargraph}$.

Each host that attaches to the JSDA network runs a message routing program, which attaches to a parent specified on the command-line.

Each service program registers itself with the local host message router. A map of the services and their locations are automatically distributed throughout the JSDA network.  Conceptually, this is similar to the exchange of route information in Border Gateway Protocol handling\cite{rfc4271}.

Client programs may invoke methods on remote hosts by instantiating a Java Object called a Proxy.  The proxy classes are automatically generated, using a GNU licensed library for decoding the Java bytecode of the servers that publish APIs to the JSDA network. This arrangement simplifies accessing remote services, illustrated with the following code:

The client connection is stateless, meaning that a java service publishing an API can be stopped on one host, started on another, and any clients attempting to access that API with a Proxy institated before the publishing change is invisible to the client.

\begin{verbatim}
// example java client code
...
import proxy.TimeProxy;
...
  TimeProxy tp = new TimeProxy();
  long[][] leapsecondstable = tp.getUTCMinusTAITable();
...
\end{verbatim}

\begin{verbatim}
# example jruby client code
...
import 'proxy.TimeProxy'
...
  tp = TimeProxy.new
  leapsecondstable = tp.getUTCMinusTAITable
...
\end{verbatim}

In the example above, the JSDA communications network takes care of discovering where the TimeServer is located and accessed via the TimeProxy object.

JSDA does not provide any resource locking mechanism within the protocol.  Some simple locking has been implemented on top of the protocol, but the commensal observing system is designed to avoid any locking of resources, (see section $\ref{sec:commensal}$).

The JSDA system was designed with a target of distributing communications between applications in a network composed of 1000-2000 individual hosts and benchmarking appears to confirm its scalability to this level.  JSDA is not intended to provide hard or soft real-time message passing.

Further services have been written to provide a subset the CORBA implementation repository (IMR) functions, such as automatic launching of services from an XML description file.  These steps were undertaken to avoid the complexities of developing CORBA services. These designs were formulated circa 2001-2002, when it was assumed the individual antenna control computers would not be capable of running a full Linux system stack, but instead would run picoJava CPUs\cite{picojava97}.

\subsection{Antennas}
Each ATA antenna is controlled by a Single Board Computer (SBC), configured with a 1GHz VIA C3 Nehemiah CPU, 512MB RAM and no permanent storage.  These computers use PXE booting to copy a Linux boot image from a central boot host, via 10BaseT ethernet over fiber. Each antenna is therefore a fully Internet Protocol addressable host.

The antenna SBC interfaces to various sub-controller boards to monitor and control the drives, receiver focus, cryogenics and low-noise amplifiers (LNA).

Each antenna runs JSDA services offering control and monitor APIs.

\subsection{Local Oscillators}

The Local Oscillators are controlled via GPIB.  All GPIB systems attached to the ATA have APIs accessable via JSDA.

\subsection{Downconverters and Digitizers}

The signal path from the antennas and into the main processing room is first converted to baseband in an RF converter board (RFCB). After this conversion step the RFCBs create four parallel outputs which are then mixed with the local oscillator signals labeled a, b, c and d (figure $\ref{fig:systemdiagram}$) providing four independent tunings from each polarization output of each antenna\cite{JWWelchIEEE09}.

These analog signals are then fed into the digitizers that are controlled via a simple text command system over TCP/IP on ethernet via Ruby scripts. 

To account for geometric changes during interferometry measurements, fringe tracking is applied at the time signals are digitized. Control of this system is accomplished via the same simple text based protocol over TCP/IP and managed with Ruby scripts.

\subsection{Correlators}

At the time of this writing, there are 2 actively used 64 input correlators for imaging purposes\cite{ATAmemo73}.  Configuration and initialization of correlators is driven by a database and support software written in Ruby on Rails and controllable by generic Ruby scripts.

Once a correlator is initialized into a particular state, it is always sampling from its inputs.  During observations, data is emitted via a Ruby based daemon as individual baseline visibility records to data catchers on a data archiving host.  These data are then written in MIRIAD format to disk\cite{miriad}.

The Master Observing Program handles the starting and stopping of data streams from the correlator host computers to the data archiver using a combination of JRuby accessing the JSDA network and Ruby Distributed Objects\cite{rubydo}.

\subsection{Beamformers}

The beamformer hardware is based on the Berkeley Emulation Engine 2 (BEE2) computing platform\cite{bee2ieee05}

Each beamformer is able to provide a sub-beam of a specific area of the current field of view, allowing multiple systems to performa analyses on the output stream. The primary application for the beamformers is to provide the signal feed to the SETI detection hardware called Prelude\cite{JWWelchIEEE09}.

The beamformers are controlled by their own host Linux instance, PXE booted from a centralized boot server and running control software written in Ruby.

\section{PROJECT SELECTION}

As of the writing of this paper the ATA accepts internal proposals with the help of a program module running under Drupal. This is a set of web-based forms and database handling routines, used to help guide the selection process of observing campaigns through the ATA Time Allocation Committee (ATAC) empowered to formulate a monthly schedule of observations.

\subsection{Drupal Module, Proposals}

Drupal was chosen in 2007 as a stable web facing platform with the ability to allow self-contained extensions called modules.  A simple module was developed to handle basic submission of proposals and tracking of proposals by observers and the ATAC.\cite{drupalsite}\supit{,}\cite{logbookproposals}

Future plans are to expand this module to allow plugable algorithms to help automatically generate a schedule that includes potential commensal observations.  Even without specific commensal observations, the master observing system (see section $\ref{sec:mop}$) will allow a secondary default observing campaign (e.g. SETI searches) to help drive the flow of observing.

This class of problem resembles a combinatorial auction\cite{combinatorialauctions}. By the time this paper is presented, we hope to have some examples of this working for the ATA.

The Drupal site is easily accessed using Ruby based calls via XML-RPC\cite{xmlrpc}.

\subsection{Observing Criteria}

Each request for observing time includes information that is required to understand how the various major components of the ATA will be utilized, such as, but not exhaustively:

\begin{itemize}
\item Targets, either in standard catalog, RA/Dec coordinates or some other algorithm for generating targets, such as mosaicing a large field.
\item Calibrators, either a standard/automated list or specific requirements
\item Number of tunings and their frequencies
\item Number of Antennas or a sensitivity target
\end{itemize}

Some observing proposals may match up in a union of resources that can be shared, i.e. antennas pointing at the same field of view, tunings that also match or available extra tunings and so on.

\section{COMMENSAL OBSERVING}
\label{sec:commensal}

After a union is found, the matching proposals are divided between "drivers seat" and "passenger" observations.  Driver seat projects are queried for their next targets, while the passenger projects may provide extra-time constraints on how long a particular field of view is monitored, e.g. the passenger observation may require more time on a calibrator than the driver, and the master observing program takes care of this.

\subsection{Master Observing Program (MOP)}
\label{sec:mop}

Until this year, all projects run on the ATA were a mixture of C-Shell scripts calling out to command-line java program clients to control the major components of the ATA.

At the time this paper is being written, we are undertaking a shift to a JRuby based Master Observing Program (MOP). JRuby provides a common layer between JSDA and the various Ruby scripts and libraries now written to control the various components of the ATA.

\begin{figure}
\begin{center}
\begin{tabular}{c}
\includegraphics[height=20cm]{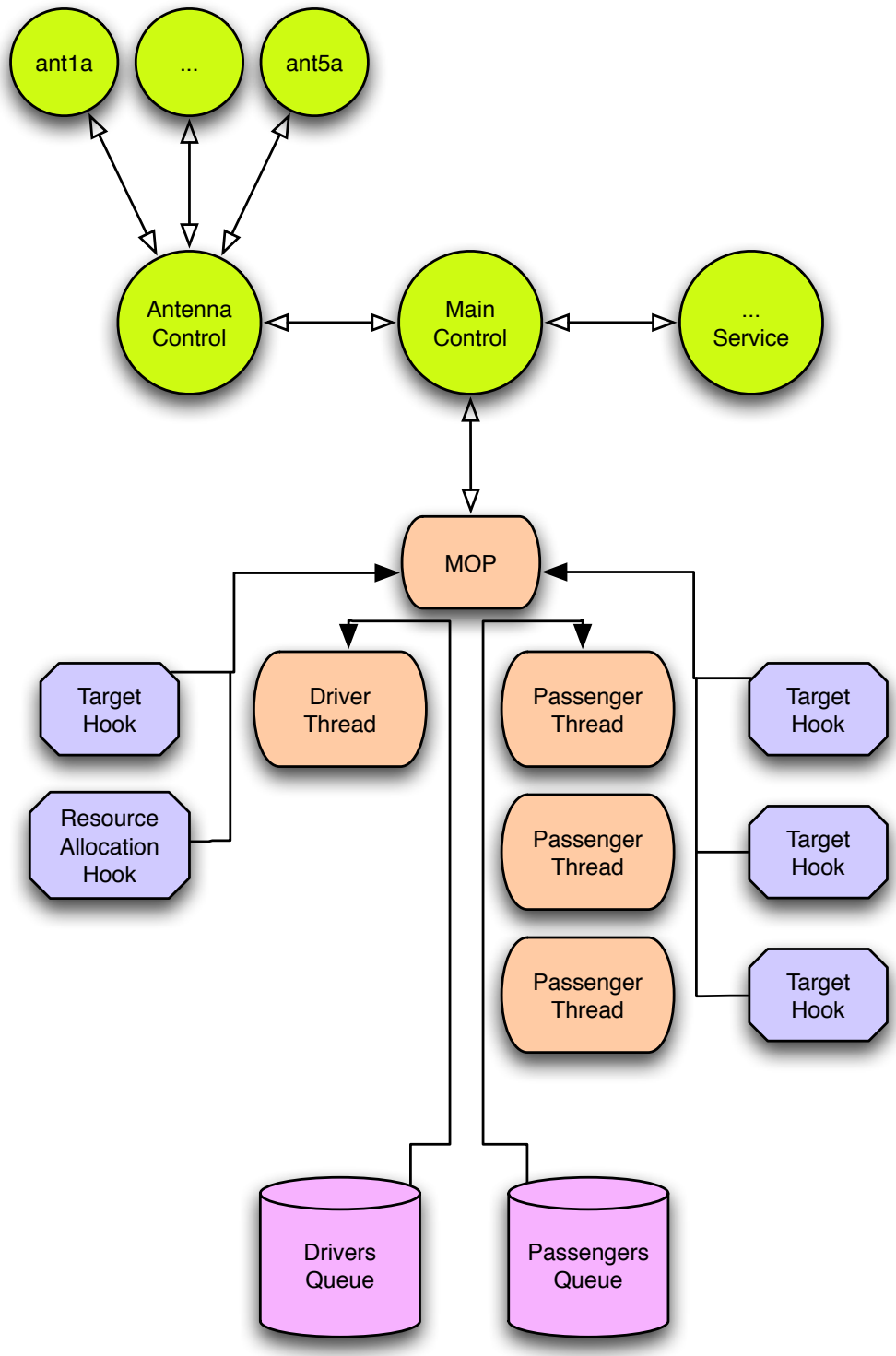}
\end{tabular}
\end{center}
\caption[Master Observing Program (MOP) interface] 
%>>>> use \label inside caption to get Fig. number with \ref{}
   { \label{fig:resources} 
By this design, locking of resources is unnecessary, as it is up to the MOP to arbitrate final commands sent to components of the ATA system while allowing complex "Astronomer Logic" operations to be controlled by observers.}
\end{figure} 

The MOP provides the primary interface between control of the system components and the guiding "Astronomer Logic" embedded in the observing scripts and various support libraries behind the decision making duties of the MOP, see figure $\ref{fig:resources}$. This also allows for continued direct access to sub-components of the ATA using existing infrastructure.

The MOP holds a queue of "driver" project observations to run.  This queue maintains state of observations that are considered to be in the "drivers" seat of the ATA as well as a separate queue of potential "passengers".  Regularly, SETI searches will be the default passenger observation.

The MOP makes calls to hook methods to both the driver and passenger observation scripts, which return information about the desired number of antennas, tuning information and where and under what name data will be archived (if it is needed).  

If there is a mismatch between what the driver and passenger scripts request, favor is given to the driver script, while a determination is made for the passenger script (again via hook methods) to determine if failure to provide a particular resource is catastrophic to the passenger script.

If so, preference is given to any other available passenger observation.

This design allows for easy integration with new and unforeseen back-end systems, such as the Flyeye experiement run on the ATA in 2008\cite{2009AAS...21460104S}. This project could be run with complex target acquisition based on a list generated by the primary investigators, while allowing the investigators to avoid having to delve into the complexities of controlling the array or sub-arrays of antennas (unless they want to).

It also allows for integration with the existing SETI observing program, but allowing the hook to call out to existing software used by the SETI team to pick out target stars within a given field of view.

Once observations are completed, automated pipeline processing software runs on targets marked as calibrators\cite{2009AAS...21460106K}.

Over the summer of 2010, work will be performed to take the automated processed data and assign grades to observations.

\newpage

\subsection{Observing Target Hooks}

Each observing program consists of either a static list of targets to observe, e.g. "M31","NGC4414" or can have a more complex list of RA/Decs or a trajectory or on the fly generation of mosaic pointings. These targets are provided to the MOP via a hook that is part of each observing program.

By default, this hook is defined in Ruby as:

\begin{verbatim}
module Observing
  class ProjectNNNN
    def targets_hook
      Observing::OBSLIST
    end
  end
end
\end{verbatim}

OBSLIST is defined in a setup file, in the Observing name space, created for each observation.

This hook is overridable, in a normal Ruby way, by creating a tag-along library for each observation, allowing observers to create complex behaviors for pointing, as they have the full power of Ruby, plus access to state information from the array itself, e.g.

\begin{verbatim}
module Observing
  class ProjectNNNN
    def targets_hook
      # generate complex pointings mosiac and return next target(s)
      # based on previous observations
    end
  end
end
\end{verbatim}

\section{CONCLUSIONS}

An often repeated concern among the astronomy community is that data analysis and instrument control software becomes overly complex and may hinder further engineering innovation.  The design presented here creates a simplified controlling mechanism that interlocks resource allocation by design, without layering that control into the underlying control libraries already implemented over the last decade for the ATA.  This creates opportunities for any observer to create rich observing behaviors and new "Astronomer Logic" without having to navigate a convoluted locking mechanism to ensure commensal observations do not interfere with each other.

Common behaviors are simple to execute while deep modifications to how observations are run are straight forward.

%%%%%Sometimes it is necessary to precede the double slash 
%%%%%by \verb|\protect| to get the desired result, 
%%%%%for example, in article titles.

%%%%%%%%%%%%%%%%%%%%%%%%%%%%%%%%%%%%%%%%%%%%%%%%%%%%%%%%%%%%%
\acknowledgments     %>>>> equivalent to \section*{ACKNOWLEDGMENTS}       

This work was supported in part by the Paul G. Allen Family Foundation under Grant 5784, the National Science Foundation under Grants 0540599 and 0540690, Nathan Myhrvold, Greg Papadopoulos, Xilinx Inc, the SETI Institute, the Unversity of California, Berkeley and other private and corporate donors.

%%%%%%%%%%%%%%%%%%%%%%%%%%%%%%%%%%%%%%%%%%%%%%%%%%%%%%%%%%%%%
%%%%% References %%%%%

\bibliography{commensal}   %>>>> bibliography data in report.bib

\begin{thebibliography}{10}

\bibitem{JWWelchIEEE09}
Welch, J. et~al., ``{The Allen Telescope Array: The First Widefield
  Panchromatic, Snapshot Radio Camera for Radio Astronomy and SETI},'' {\em
  Proc. IEEE}~{\bf 97}(8),  1438--1447 (2009).

\bibitem{ATAmemo30}
Wright, M., ``Astronomical imaging with the {ATA} - {III},'' (2003).
\newblock http://ral.berkeley.edu/ata/memos/memo30.pdf.

\bibitem{ruby}
Matsumoto, Y. et~al., ``{Ruby Programming Language}.''
\newblock http://ruby-lang.org.

\bibitem{ror}
Hansson, D.~H. et~al., ``{Ruby on Rails}.''
\newblock http://rubyonrails.org/.

\bibitem{jruby}
Nutter, C., Enobo, T., et~al., ``{JRuby, Pure-Java Implementation of the Ruby
  Programming Language}.''
\newblock http://jruby.org/.

\bibitem{omgcorba}
{Object Management Group}, ``{Common Object Request Broker Architecture},''
\newblock http://www.omg.org/spec/CORBA/.

\bibitem{rfc4271}
Rekhter, Y., Li, T., and Hares, S., ``{A Border Gateway Protocol 4 (BGP-4)}.''
  RFC 4271 (Draft Standard) (Jan. 2006).

\bibitem{picojava97}
{O'Connor}, J.~M. and Tremblay, M., ``{picoJava-I}: The java virtual machine in
  hardware,'' {\em IEEE Micro}~{\bf 17}(2),  45–53.
\newblock doi:10.1109/40.592314.

\bibitem{ATAmemo73}
Urry, W., Wright, M., Dexter, M., and MacMahon, D., ``{ATA} memo 73: The {ATA}
  correlator,'' (2007).
\newblock http://ral.berkeley.edu/ata/memos/memo73.pdf.

\bibitem{miriad}
``{MIRIAD}: {M}ultichannel {I}mage {R}econstruction {I}mage {A}nalysis and
  {D}isplay,''
\newblock http://bima.astro.umd.edu/miriad/.

\bibitem{rubydo}
Seki, M., ``{Distributed Ruby: dRuby},'' (2009).
\newblock http://www.ruby-doc.org/stdlib/libdoc/drb/rdoc/index.html.

\bibitem{bee2ieee05}
Chang, C., Wawrzynek, J., and Brodersen, R.~W., ``{BEE2}: A high-end
  reconﬁgurable computing system,'' {\em IEEE Design and Test of
  Computers}~{\bf 22}(2),  114–125.
\newblock doi:10.1109/MDT.2005.30.

\bibitem{drupalsite}
``Drupal,''
\newblock http://drupal.org.

\bibitem{logbookproposals}
``{HCRO LogBook, Proposals},''
\newblock http://log.hcro.org/ata-proposals.

\bibitem{combinatorialauctions}
Cramton, P., Shoham, Y., and Steinberg, R., ``Combinatorial auctions,'' (2006).
\newblock ISBN 0-262-03342-9.

\bibitem{xmlrpc}
``{XML-RPC Specification}.''
\newblock http://www.xmlrpc.com/.

\bibitem{2009AAS...21460104S}
{Siemion}, A., {Bower}, G., {Cordes}, J., {Foster}, G., {Mallard}, W.,
  {McMahon}, P., {van Leeuwen}, J., {Wagner}, M., {Werthimer}, D., and {Allen
  Telescope Array Team}, ``{Results from the Allen Telescope Array: The ATA
  Fly's Eye Transient Search},'' in [{\em American Astronomical Society Meeting
  Abstracts}{\nolinebreak\hspace{0.1em}]},  {\em American Astronomical Society
  Meeting Abstracts} {\bf 214},  601.04--+ (Dec. 2009).

\bibitem{2009AAS...21460106K}
{Keating}, G., {Barott}, J., and {Allen Telescope Array Team}, ``{{Results from
  the Allen Telescope Array: Real-Time Imaging}},'' in [{\em American
  Astronomical Society Meeting Abstracts}{\nolinebreak\hspace{0.1em}]},  {\em
  American Astronomical Society Meeting Abstracts} {\bf 214},  601.06--+ (Dec.
  2009).

\end{thebibliography}
\bibliographystyle{spiebib}   %>>>> makes bibtex use spiebib.bst

\end{document}